\documentclass{appolb}
\usepackage{epsfig}
\usepackage{amssymb}
\usepackage{amsmath}
\newcommand{\als}{\alpha_s}
\newcommand{\epem}{e^+e^-}
\newcommand{\pom}{I\negmedspace P}
\begin{document}
%\date{\today}
\pagestyle{plain}
%% uncomment the following line to get equations numbered by (sec.num)
%\eqsec
\newcount\eLiNe\eLiNe=\inputlineno\advance\eLiNe by -1
\title{MULTIPARTICLE DYNAMICS 2004}
\author{Wolfgang OCHS
\address{Max-Planck-Institut f\"ur Physik, Werner-Heisenberg-Institut\\
F\"ohringer Ring 6, D-80805~M\"unchen, Germany}}
\maketitle

\begin{abstract}
We summarize results presented at this conference
with special emphasis on hard processes 
with jets and heavy quarks, soft particle production,
small x structure functions and diffraction as well as heavy ion collisions
and quark gluon plasma.
\end{abstract}

\section{Introduction}
In high energy collisions of leptons,
hadrons and nuclei
we observe the production of many particles, mainly hadrons.
The ultimate goal in multiparticle production 
studies is the explanation of the hadronic phenomena within QCD, 
the basic theory of the strong interactions, along with the other interactions 
of the standard model and possibly beyond.

A basic problem in the QCD study of multiparticle production is the matching
of parton and hadron dynamics relevant in the respective 
weak and strong coupling regimes of QCD. A class of inclusive observables
in hard collisions can be computed perturbatively
in terms of the running coupling constant $\alpha_s$ thanks to the celebrated 
asymptotic freedom \cite{gwp}.  
In practice, it is often required to include some additional non-perturbative
input from other sources (\eg Parton Distribution Functions).
Hadronic phenomena can be systematically analysed within lattice 
gauge theory in sufficiently simple problems. The description of
genuine multi-hadron production requires, in addition to the
hard QCD part (shower calculus), some phenomenological approaches: 
specific hadronization models or such simple ideas as 
``parton hadron duality''. At present,
the systematic approach to multi-hadron production 
based entirely on QCD remains a dream. Is it just a problem of complexity
or do we need a fundamentally new insight?

In this talk 
we concentrate on the four topics mentioned in the abstract 
which represent different variants 
of the interplay of hard and soft interactions and 
we emphazise shortly other topics.
I am sorry for the incomplete coverage of the many interesting presentations
at this conference.

\section{Hard processes with jets and heavy quarks}
Results are reported from TEVATRON, HERA,
RHIC and LEP accelerators. Of 
central importance is the comparison with
fixed order perturbation theory 
to test the universality of the coupling constant $\alpha_s(Q^2)$
in all processes.
Here the goal is to improve the
accuracy of the calculation, in particular by the extension 
beyond Next-to-Leading-Order
accuracy, and to compute within the QCD framework 
observables of higher complexity.  
An important goal is also the discovery of new physics either through the
deviation of experimental results from the precision calculations
or through better understanding of background processes.

{\it Top quark production (D. Bauer)} has been studied
at the TEVATRON $p\bar p$ collider
where in the new RUN II the cms energy is increased from 1.8 to 1.96 TeV.
Top quarks at these energies are produced primarily in pairs and they decay
through $t\to Wb$. The various decay channels for the $W$ have been analysed
by both experiments and consistent results have been obtained
in eight channels, all compatible with the 
% \begin{equation}
%\sigma_{t\bar t}=7.7^{+3.4+4.7}_{-3.3-3.5}\pm 0.4 \text{pb  (D0)}
%\sigma_{t\bar t}=7.9\pm 2.5^{+4.7}_{-2.3} \text{pb  (CDF)},
%\end{equation}
%with statistical, systematic (and lumi) errors.
 theoretical computations beyond NLO of 
$\sigma_{top}=6.77\pm0.42$ pb at $m_t=175$ GeV \cite{kv}.
A new value for the top quark mass has been presented by D0 \cite{d0top}
from a reanalysis of their earlier RUN I data: $m_t=180.1\pm3.6
({\rm stat.})\pm 3.9 ({\rm syst.)}$ GeV with considerably reduced errors
and this result increases the world average by 4 GeV to
$m_t=178\pm4.3$ GeV. The best RUN II CDF result so far obtained is
$m_t=177.8^{+4.5}_{-5.0}\pm6.2 $ GeV. 
No single top production has been observed and
$\sigma_t<8.5$ pb at 95\% CL (RUN II, CDF).

 {\it Single top production
(S.D. Ellis)}.
%It can proceed through spacelike and timelike $W$ exchange. 
%process ($q\bar q\to W\to \bar t b$) with an expected $\sigma\sim 0.88 $ pb
%or a space like process ( $bq \to tq$) with $\sigma\sim 1.98 $ pb at cms
%energy $\sqrt{s}=1.96$ TeV. 
The observation of this process would determine
the CKM matrix element $V_{tb}$ and be
important in other searches (Higgs, new particles,
including extra scalar bosons or gauge bosons, new quarks). The expected
rates are still below the presently achieved sensitivity. Therefore some
strategies to obtain an improved signal/background ratio  are proposed,
especially by using the ``signed rapidity'' variable
which takes into account the fact that processes with $W$'s  
are not separately C or P invariant.
\begin{figure}[tb]
\unitlength1cm
 \unitlength1cm
 \begin{center}
% \begin{minipage}[tbh]{9.5cm}
           \mbox{\epsfig{file=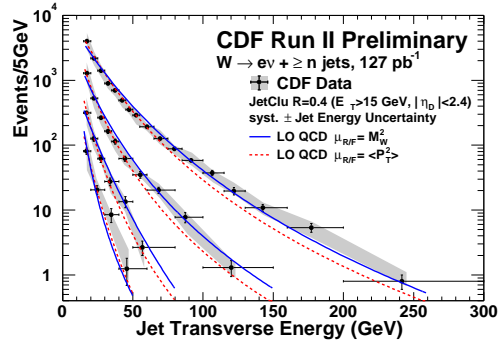,width=7cm}}
% \end{minipage}
\end{center}
\caption{\label{wjet}
 Distribution of the jet E$_{T}$ for the jet of lowest energy in the
 $W +\geq n$ jet sample, 
%for $n=1\ldots 4$ from top downwards, 
for $n$=1 (top) to $n$=4 (down),
in comparison with LO QCD calculations.
}
\label{fig:jets}
\end{figure}

{\it Jet Production at the TEVATRON (L. Sawyer)} 
is measured now at the higher cms energy by CDF and D0. Single inclusive
jets are studied %in the rapidity range $|\eta|<2.4$
up to transverse momenta $ p_T\sim 550$ GeV, that is 
150 GeV higher than in RUN I and
di-jet masses up to $M_{ij}\sim
1400$ GeV, also the azimuthal decorrelation of the two jet events 
which is an effect of ${\cal O}(\alpha_s)^3$ is measured. These results
are found in an 
overall consistency with the
NLO QCD predictions \cite{eks,jetrad}  
%existing PDF's (CTEQ6M \cite{cteq6m}) 
in the extended regime of
energy and transverse momentum within the systematic errors which are
dominated by the energy scale uncertainty.  Only in extreme 
kinematic regions of the
azimuthal decorrelation deviations appear. \\

{\it Production of W,Z,$\gamma$ + jets (A. Cruz)} 
allows further tests of pQCD at
large momentum transfers using different techniques. The inclusive
W production cross section rises from $2.38\pm 0.24$ nb at 1.8 TeV to
$2.64\pm 0.18$ nb at 1.96 TeV and is well described by the NNLO QCD results
of 2.5 and 2.73 nb respectively. 
For the production of W together with n jets the distribution of
 the jet transverse energies have been measured.
There is good agreement with a calculation (ALPGEN
\cite{alpgen}) based on leading order QCD Matrix Element combined with
parton shower development (HERWIG \cite{herwig}), 
see Fig. \ref{fig:jets}. Here the experimental
uncertainties from the jet energy scale are comparable in size to the
theoretical scale uncertainties.
{\it Central di-photon production}
has been measured up to masses of about 30 GeV.
The NLO QCD calculations (DIPHOX \cite{diphox}) 
involving the $q\bar q$ annihilation and quark loop
diagram for $gg\to \gamma\gamma$
describe the data well in absolute normalization. The 
{\it associated production
of photon with heavy flavours (c,b)} has been obtained as well
for photon energies up to 60 GeV. It agrees
with a LO calculation (Pythia \cite{pythia}).

In summary for TEVATRON, no serious disagreement of the new data with QCD
expectations at different levels of accuracy for the observables 
of different levels of complexity are
met and so, at the same time, we do not 
obtain any signal of a new physics. The
other problems discussed here concern the role of the ``underlying
event'' ({\it R.D. Field}, see below) 
and the influence of the selection of jets by a
particular algorithm (``cone'' vs.``$K_T$'') ({\it Andrieu}).

{\it Monte Carlo generators, (L. L\"onnblad)} are being developed to generate 
parton final states beyond LO for present
accelerators but also for the LHC. For
problems like production of multi-jet or W+jet events a matching of the NLO
matrix element and parton shower has been achieved; also double scattering
effects are being considered.

{\it Fragmentation functions (S. Kretzer)} $D_{q,g}^{h^\pm}(z,Q^2)$ 
are compared in $pp$ and $e^+e^-$ collisions as a tests the collinear
factorization approach of QCD at NLO together with
  %$Q^2>\mu^2>\Lambda^2$ 
universality. New data obtained at RHIC on $pp\to \pi
X$ are well predicted using the previous results \cite{kkp,kretzer} from
$e^+e^-$ collisions 
on $D_{q,g}^{h^\pm}$ at the factorization scale and subsequent DGLAP evolution.

{\it Jet production at HERA (C. Glasman)} 
has now been measured down to $x\simeq 10^{-4}$
for momentum transfers $Q^2$ larger 
than a few GeV$^2$.
The $Q^2$ dependence of of two- and three-jet cross sections in NC DIS
up to $Q^2 \sim 5000$ GeV$^2$, obtained by ZEUS, is 
found in good agreement with
NLO (${\cal O}(\alpha_s^2)$ and ${\cal O}(\alpha_s^3$)) predictions.
The ratio of both results provides an accurate test of the theory 
and in particular a precise determination of the coupling constant
$\alpha_s(M_Z)=0.1179\pm0.0013$  
(stat.)$^{+0.0028}_{-0.0046}$ (exp.) 
which by itself compares well with the current world average
$\alpha_s(M_Z)=0.1182\pm 0.0027$ \cite{bethke} but 
there is still a large theoretical error of ($+0.0061, -0.0047$).

Of special interest, also in view of further applications to small $x$
physics, is a test of the validty of the DGLAP approximation to the $Q^2$
evolution of structure functions. 
%This approximation corresponds to the
%summation of the $\log Q^2$ power series and according to the analysis of
%Feynman diagrams to the strong $k_T$-ordering of the partons emitted before%
%hard scattering. This is no longer appropriate if, for small $x$, terms%
%proportional to $\alpha_s \log 1/x$ become important. Some evidence has been
%presented for such a breakdown.
%
If jets are produced ``forward'' (in direction of incoming proton) and have
transverse energies $(E_T^{jet})^2$ of order $Q^2$ then the kinematic
configuration is not in favour of intermediate gluon emission with 
strong $k_T$ ordering as is typical for DGLAP
evolution and deviations from DGLAP based predictions are expected
\cite{mueller}.
 Indeed, if jets are selected by the H1 collaboration 
with $0.5< (E_T^{jet})^2/Q^2<5$ and $x<0.004$, then 
the ``direct'' NLO QCD predictions  (DISENT \cite{disent}), 
\ie without photon structure,  are too low by about a factor 3 for very
small $x<0.001$. % while there is agreement for $x\gtrsim0.002$. 
Considerable improvement, although not full agreement, 
is obtained if a photon structure is included in the DGLAP calculation. If
one of the two scales $E_T$ or $Q^2$ is large compared to the other one the
DGLAP factorization approach works best. Whereas work continues to develop the 
approximations beyond DGLAP, there is no serious conflict with pQCD at 
the fundamental level.

{\it Spin physics}.  
The spin program at RHIC {(\it E. Sichtermann)}  aims at a measurement of
of hard and soft processes with polarised protons. 
The single transverse spin asymmetry $A_N$ of forward $\pi^0$ production 
has been observed at FNAL at  $\sqrt{s}=20$ and the first results from RHIC
show that it persists at $\sqrt{s}=200$ GeV. 
%The double 
%longitudinal spin asymmetry $A_{LL}$ is found small at mid-rapidity.
New results are also presented  from the HERMES experiment at DESY 
({\it I. Gregor}) 
on the study of the transverse single-spin asymmetries in semi-inclusive pion 
production in DIS.  This allows the 
determination of the protons ``transversity'' distribution, 
which represents the degree
to which the quarks are polarised along the proton spin transversely 
polarized to the virtual photon. 

\section{Soft particle production in hard collisions}

The classical applications of pQCD concern observables for hard processes
where the hadrons are either summed over or collected into jets
so that fixed order perturbation theory can be applied, as in 
the previous section. 
A further development, surprisingly successful, is
the application of pQCD to observables calculated from the individual momenta
of hadrons in the final state directly and in general involves a resummation 
of the perturbation theory. 
One has to include some assumptions on the transition from
partons to hadrons and eventually on some non-perturbative aspects of colour
confinement. 
There are final state parton observables, like event shapes, 
which are infrared and
collinear safe, i.e.  their
values do not change if a collinear or soft gluon is added.
They are less sensitive to soft hadronization effects. 
More sensitive are multiplicity observables, such as particle flows inside 
jets or between jets which are not infrared safe. 
These observables are a testing ground for ``parton
hadron duality'' ideas
(review by {\it Yu. Dokshitzer}). 

\subsection{Infrared and collinear safe observables}
{\it Event shapes (Yu. Dokshitzer)}. 
These observables describe global
properties of hadronic final states, in $e^+e^-$ annihilation, for example,
one defines ``thrust'', ``jet mass'', 
``broadening'' and others.
The analysis of these observables in perturbation theory
(an asymptotic expansion) 
leads to a description which combines perturbative and non-perturbative 
aspects,
the latter ones represented by a power correction $\propto(1/Q)$ \cite{dwm}.
%For the mean values one finds generically
%
%\begin{equation}
%{\cal V} =A\als+B\als^2+c_\nu\frac{2C_F}{\pi}\frac{\mu_I}{Q}
%(\alpha_0-\langle \als^{PT}\rangle_{\mu_I})
%\end{equation}
%with NLO terms $A,B$ and the process dependent factor 
%$c_{{\cal V}}$. 
This term 
%which represents the higher order corrections, 
involves an integral 
over gluon emissions at small transverse momenta $k_T$
where the coupling constant $\alpha_s(k_T^2)$ is ill defined. 
It is assumed that this integral is finite and
 - as the coupling itself - universal for the different observables.
One therefore introduces the parameter 
$ \alpha_0=\frac{1}{\mu_I}\int_0^{\mu_I} dk_T \alpha_s(k_T^2)  $
at the matching scale $\mu_I$
to describe the influence from the soft region \cite{alpha0}.
This result, strictly obtained for partons, is then applied to the 
experimental hadronic observables assuming a duality  between both
descriptions. The calculation has been extended to the differential
distribution of shape observables where the non-perturbative effects 
can shift or squeeze
the perturbative spectra by an amount given by $1/Q$.
By now, the fits to event shapes in $\epem$  with two parameters 
provide a competitive determination of $\alpha_s$
with an error of $\sim 8\%$ and the universality of the non-perturbative
parameter
$\alpha_0\sim 0.5$ at $\mu_I=2$ GeV is confirmed within $\sim 15\%$.

{\it  Angularities (G.Sterman)}: 
A new class of event shapes allows specific tests for
the non-perturbative corrections. From the angles $\theta_i$ of the particles 
to the thrust axis and energies $E_i$ one constructs the quantity \cite{bks}
\begin{equation}
 \tau_a=\frac{1}{Q} \sum_i E_i(\sin \theta_i)^a(1-|\cos\theta_i|)^{1-a}
\end{equation}
which interpolates between thrust ($a=0)$ and broadening ($a=1$).
Again, one can separate a contribution from soft gluon emission
which is then represented by a non-perturbative ``shape function''
$f_{a,NP}$. It represents corrections of all higher orders in $\Lambda/Q$    
which should be more appropriate near the collinear limit. This represents
a generalization of the correction $\alpha_0/Q$ above, 
which corresponds to a shift of the distribution.
%The observed distribution in
%$\tau_a$ is then given by the convolution
%$\sigma(\tau_a,Q)=\int_0^{\tau_aQ}d\xi f_{a,NP}(\xi)\sigma_PT(\tau_aQ-\xi,Q)$
%This can be viewed as a generalization of the correction term $alpha_0/Q$
Once determined at a particular energy one obtains predictions
for other energies. 
%The simple $a$ dependence of the shape function allows a
%test on the assumed rapidity independence of the non-perturbative
%dynamics.\\ 

{\it Interjet radiation: non-global log's}. 
In DIS and
hadron-hadron collisions one is led to consider gluon radiation in part of the
phase space excluding a region around the beam direction, so energy flow or
event shapes are non-global.  Such observables obtain contributions from
multi-soft emissions which lead to ``non-global log's'', single
logarithmically enhanced contributions \cite{ds}. The difficulty is that the
number of jets is not fixed. This problem can be tamed by construction of a
correlation of the energy flow with an event shape which fixes the number of
jets \cite{bks,dm}. 
%Results are obtained at leading logarithm in energy flow and NLL
%in the event shape.

{\it Automated resummation (G. Zanderighi)}: 
In order to facilitate the
calculation of new shape observables, in particular for $p\bar p$ and DIS,
a program (CAESAR \cite{caesar}) 
has been developed which,
for a class of observables,  yields resummed results in NLL order by
combining analytical and numerical methods. The limitations concern certain
properties in the infrared and collinear limit of the emission
(``recursively IRC-safe'') and the dependence on transverse momentum
(``continuously global''). 
Old results have been reproduced, new observables derived,
for example, the distribution of global transverse thrust
$
T_\perp = \frac{1}{E_T}\ \text{max}_{\vec{n}_T} \sum_i|\vec{p}_{t_i}\vec{n}_T|
$
constructed from the transverse momenta with respect to the beam axis. 
Such calculations open up
 the possibility to considerably extend the kinematic range of these QCD
studies towards the higher energies of TEVATRON and LHC.

\subsection{Multiplicities, particle flows}
These observables are not infrared safe: emission of a soft gluon would
increase the multiplicity by +1. Finite perturbative results for the parton
cascade can be obtained by introducing a cut-off $k_T\geq Q_0 $. For $Q_0\gg
\Lambda$ the partons represent jets and $Q_0$ can be viewed as jet
resolution in the sense of the ``$k_T$-algoritm'', one can also take the
small $Q_0 \gtrsim \Lambda$ and compare the resulting cascade directly with
the hadronic final state in the sense of a duality picture
(``Local Parton Hadron Duality'' \cite{dkmtbook}), then $Q_0$ is a
non-perturbative parameter.

{\it Multiplicities in quark and gluon jets at LEP (K. Hamacher) and 
TEVATRON (A. Pronko)}: At LEP the 
multiplicity of gluon jets is determined from 3 jet events after subtraction of 
 2 jet events at a reduced scale (DELPHI) or from 3 jet events 
using a boost algorithm (OPAL \cite{opalmult}).
The quark jet 
multiplicity is found directly from the total $\epem$ multiplicity. 
Theoretical results are obtained from resummed perturbation theory.
One approach is based on
coupled evolution equations of quark and gluon jets in Modified Leading
Logarithmic Approximation \cite{dkmtbook} which includes fully the 
$\sqrt{\als}$-correction up to NLL order.
These calculations reproduce the multiplicity rise with energy.
%The global event multiplicity follows the MLLA prediction
%$\bar N \propto \als(Q)^C\exp(C2/\sqrt{\als(Q)})$ up to the highest energies
%Q=200$ GeV. 
The ratio $r=N_g/N_q$ obtains large corrections beyond MLLA and is reduced 
from the asymptotic value $r=C_A/C_F=9/4$ to $r=1.7$ in  3NLLO
\cite{dremin} and to $r=1.5$, observed at LEP, 
in the numeric solution \cite{lo}
where the only parameters $\Lambda,Q_0$ are fit by the total 
$\epem$ multiplicity. 
Another calculation is based on the colour dipole model which treats the
evolution of dipoles in NLL approximation and includes recoil effects
\cite{dipole}. It describes the data well using an additional 
non-perturbative parameter. 

The CDF Collaboration has separated quark and gluon jets by
analysing di-jet and $\gamma$ + jet events with known jet compositions. The
multiplicity data are found generally well consistent with $\epem$ results.
%especially the gluon jet results agree well with OPAL in the range $10\leq
%Q\leq 40$ GeV whereas the quark jet multiplicities are a bit lower.
The multiplicities reach the higher energies $Q\sim 300$ GeV where
they also follow the 3NLLA expectations.

{\it Particles with low momenta in jets  
(A. Pronko)}, measured by CDF, show the so-called hump-backed plateau
 in the variable $\xi=\log(1/x)$ 
\cite{dkmtbook} with the suppression of the low energy, large $\xi$
particles because of soft gluon coherence. Results on the ratio $r(\xi)$ of
these spectra for gluon over quark jets for $\xi>3$ approach ratios of 
$r(\xi)\sim 1.8\pm 0.2$ again in good agreement with OPAL results.
Although larger than for the full jet result ($r\sim 1.5$)
it is still below $r(\xi)=C_A/C_F$ expected in \cite{klo} for this limit
from the dominance of the primary gluon emission. This discrepency is likely
due to the difficulty to obtain "pure" gluon jets, rather, soft
particles are emitted from all participating jets of the event.
This problem is avoided in

{\it Soft particle emission in 3-jet events in $\epem$ (Hamacher)}.
The particle multiplicity $N_3$ in a cone
perpendicular to the production plane is studied as function 
of the inter-jet angles $\Theta_{ij}$. The gluon radiation into this cone
coherently emitted from the $q\bar qg$ ``antenna'', 
normalized by a corresponding multiplicity $N_2$ in 2-jet events,
is given by the simple expression \cite{klo}
\begin{equation}
\frac{N_3}{N_2}
  =\frac{C_A}{C_F}r_t
  = \frac{1}{4} \frac{C_A}{C_F}\left[(1\!-\!\cos\Theta_{qg})+
(1\!-\!\cos\Theta_{\bar q g})-\frac{1}{N_C^2}(1\!-\!\cos\Theta_{q\bar q})\right]
\label{r32}
\end{equation}
The first two leading terms represent the dipoles along the $qg$ 
directions, in close analogy to QED electric dipoles, except for 
the colour factors.
The formula interpolates for aligned partons
 between a colour triplet antenna 
($q$ against $qg$) and a colour octet antenna ($q\bar q$ against $g$)
with the intensity higher by $C_A/C_F$.
\begin{figure}[bt]
\unitlength1cm
 \unitlength1cm
 \begin{center}
% \begin{minipage}[tbh]{7.5cm}
           \mbox{\epsfig{file=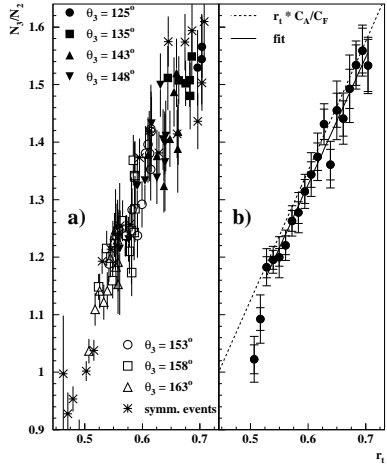,width=5cm}}
% \end{minipage}
\end{center}
\caption{\label{f:cone}
Multiplicity ratio $N_3/N_2$ in cones of $30^{\circ}$ opening angle
as function of $r_t$ in  Eq.{\protect (\ref{r32})}:
a) for different inter-jet angles $\theta_3$;
b) averaged over $\theta_3$, 
the dashed line is the expectation Eq.{\protect (\ref{r32})}
with slope $C_A/C_F$,
the full line is a fit (DELPHI \protect\cite{delphiperp}). 
}
\end{figure}

DELPHI has measured the ratio $N_3/N_2$ against the variable
$r_t$ (see Fig.~\ref{f:cone}) which shows the scaling behaviour in the angles 
implicit in (\ref{r32}) and the linear dependence
with slope $ 2.211\pm0.014\ (stat.) \pm 0.053\ (syst.)$
well consistent with the expected slope $C_A/C_F$ \cite{delphiperp}.
The data are sensitive to the small term $\propto 1/N_C^2$ which
corresponds to a negative interference not accessible through 
purely probabilistic jet algorithms. It is remarkable that the perturbative
calculations describe production of particles with such 
low momenta (few 100 MeV above threshold) at low multiplicity.

Similar transverse effects are expected for $\gamma p$ collisions
($q\bar q$ antenna in direct, $gg$ in resolved processes)
and $p\bar p$ collisions (low transverse radiation in direct $\gamma$ or $W$ 
production, high radiation in gluon jet production).

{\it Underlying event in $p\bar p$ collisions (R.D. Field)}:
Soft particle production in central rapidity perpendicular to the 
high $p_T$ trigger
jet direction increases rapidly with the jet transverse momentum
 from low $p_T$ (like minimum bias) up to about 7 GeV and saturates beyond.
The same is observed for the 
transverse momentum sum in back-to-back jet events.
If one triggers in addition on a particle in transverse direction one observes 
the ``birth'' of a third jet in the same and possibly even
on a fourth jet in the opposite direction. This conclusion is derived from the good agreement with
the PYTHIA MC which includes multiple parton collisions whereas HERWIG without 
this addition is in a less good agreement. It will be interesting in future 
studies to clarify the role of multi-parton scattering and also to investigate
the possible reduction of the transverse particle production in direct 
$\gamma$ and $W$ production processes as expected for the perturbative 
gluon radiation mechanism emphasized in the previous paragraph 
for $\epem$ collisions. 

\section{Small $x$ structure functions and diffraction}
There is an old expectation for the parton density at small $x$ to
``saturate'' \cite{glr}, i.e. to become
so high that a limiting behaviour related to the
finite proton (or nuclear) size is reached.
Ultimately, one expects a transition into a 
strong coupling regime not accessible to perturbative treatment.
With the new data from HERA and RHIC this debate enters a new round.
The  ``Pomeron'' which describes diffractive 
processes with vacuum quantum number exchange
is treated as composite object with a partonic sub-structure.

\subsection{Deep Inelastic Scattering  and parton saturation}

In the standard perturbative treatment of DIS (DGLAP 1972-1977)
the photon interacts with the proton through exchange of a single parton 
ladder which leads to a linear evolution equation of the parton densities 
in $Q^2$.
 With decreasing Bjorken $x$ there is the possibility of multiple interaction
of photon and proton through exchange of two or more parton ladders
as described by GLR \cite{glr} (1983). This happens for sufficiently 
large parton overlap probability
$W(x,Q^2)$, given by
the ratio of the parton parton cross section at scale $Q^2$,  
$\hat\sigma\sim \frac{\als(Q^2)}{Q^2}$,
to the mean distance of two partons in the proton, 
$\Delta\vec b^2\sim \frac{F(x,Q^2)}{\pi R^2}$ for proton radius $R$ and 
parton density $F(x,Q^2)\sim xG(x,Q^2)$.
While for  $W\ll 1$ the DGLAP approximation 
is appropriate, for $W\sim \als$ the non-linear recombination processes 
set in 
and, ultimately, at $W=1$ saturation is reached, i.e. a full overlap of 
partons in the proton which is beyond perturbation theory.
 This limit defines 
the characteristic saturation scale $Q_s(x)$ from
\begin{equation}
 \frac{xG(x,Q^2_s)}{\pi R^2}\sim \frac{Q_s(x)^2}{\als(Q_s^2)}
\label{qsat}  
\end{equation}
For small $\als$ this corresponds to a state of high gluon density. 

The theoretical analysis starts from the 
dipole picture \cite{disdipole}, formulated in space time,
from which the $\gamma^*p$ total cross section can be computed as 
\begin{equation}
\sigma^{\gamma^*p}_{T,L}(x,Q^2)= \int d^2r\int dz \hat\sigma_{\rm dipole}
    (\vec r,x)|\psi^\gamma_{T,L} (\vec r,z,Q^2)|^2
\label{gbwmodel}
\end{equation} 
where $\psi^\gamma$ is the wave function of the virtual photon
splitting into a $q\bar q$ dipole, $z$
the photon longitudinal momentum fraction of the quark and $r$ the 
transverse size of the dipole. 

Different approaches are used for the dipole cross section.
The scattering process can be studied in the proton rest frame
and one considers higher orders to the $q\bar q$ wave function. 
A non-linear evolution equation for the 
dipole-proton scattering amplitude has been given by Balitsky and 
by Kovchegov \cite{bk}.  
A complementary approach treats 
the gluons at small $x$ in an effective field theory in a frame with 
a low momentum photon and an 
energetic proton where the partonic motions in the proton 
are largely frozen in. 
The state of these high density gluons
is also called ``Colour Glass Condensate'' (CGC) \cite{cgc}.
This new kinematic regime of perturbative QCD at high density at the border
to a non-perturbative confinement regime is 
under intense investigation and is important 
also for heavy ion collisions where the gluonic state at
high density appears initially.
Status and applications of the theory of saturation 
and the CGC are reviewed by 
{\it J. Bartels} and {\it E.G. Ferreiro}.

\subsection{Evidence for saturation at HERA?}

According to the above outline one may reach the saturation region 
in DIS either by decreasing $x$ at fixed $Q^2$, i.e. by increasing 
the parton density
 or by decreasing $Q^2$ at fixed $x$, i.e. by increasing the parton 
transverse ``size''.
In the first case one observes that 
the structure function $F_2$ can be well fitted
by the NLO QCD in the DGLAP approach for $Q^2\gtrsim 2$ GeV$^2$ where for 
$x<0.01$ one finds an $x$ independent  slope 
$\lambda(Q^2)=-(\partial \ln F_2/\partial\ln x)_{Q^2}$
which depends linearly on $Q^2$. For smaller $Q^2$
the applicability of the perturbative calculations may be questioned,
so there is no direct evidence for saturation in the perturbative regime
from this point of view.

On the other hand, a new regime appears
in the second case when $Q^2$ is decreased at fixed $x$. In this case
it is observed that the  $\lambda$ slope saturates  
for $Q^2\lesssim 1$ GeV$^2$ ({\it talk by E. Elsen}). This
region of low $Q^2$ is included in the models of saturation 
which combine perturbative and 
non-perturbative aspects. Some essential features of this approach 
are contained already in the model by
Golec-Biernat and W\"usthoff \cite{gbw}. Here the DIS cross section is
obtained in the dipole picture (\ref{gbwmodel}) with a simple ansatz
for the dipole cross section $\sigma_{\rm dipole}(r^2 Q_s^2(x))$ 
to depend only on the particular combination of $r$ and $x$ and with
$Q_s^2(x)\sim x^{-\lambda}$ the saturation scale.
The cross section 
 is calculated perturbatively for
small distances ($\sigma_{\rm dipole}\sim r^2$) whereas at large distances
a simple Gaussian form has been adopted with saturation built in
($\sigma_{\rm dipole}\to \sigma_0$). 
In this way the full $Q^2$ range becomes accessible
in the model.

An important prediction of the model
is the geometrical scaling \cite{geomscal}  
which states that the cross section
$\sigma^{\gamma^*p}_{tot}(x,Q^2)=f(\tau)$ depends only on the quantity
$\tau=Q^2/Q^2_s(x)$. This scaling property is well satisfied.\footnote{An 
alternative scaling law
has been derived within the framework of a generalized 
vector-dominance model \protect\cite{schild}.} Another successful prediction
of the model is the near constancy of the ratio of the diffractive 
over the full cross section $F_2^{diff}/F_2^{tot}$ under variation of
the total hadronic mass $W$. 

Recent studies have removed some shortcomings of the model, especially 
the phenomenological parametrizations for large dipole sizes $r$. 
The large $Q^2$ 
behaviour can be recovered by a smooth matching to the DGLAP evolution 
\cite{bgk}. The solutions of the BK equation determine the behaviour for
large $r$ and together with the DGLAP behaviour at small $r$
a good description of HERA data is obtained \cite{gllm, iim}: the geometrical 
scaling in the saturation domain, the transition between
the hard and soft photon region for the slope $\lambda$
and the $x$-dependence of the saturation scale $Q_s^2$.
%The radius in the proton which confines the gluons below the saturation 
%scale is found as
%$R\simeq 0.3 fm$ \cite{gllm}, smaller than the proton's charge radius.
These successes in the description of data involving the scale $Q_s$ can be 
taken as indirect evidence for the onset of saturation. 

%The gluon density could be accessible in an extended
%kinematical range at small $x$ and with higher accuracy 
%by measuring also the longitudinal 
%structure function $F_L$ at HERA with improved instumentation
%but else from RHIC or LHC ({\it E. Elsen}).

\subsection{Hard diffraction in DIS}
Events with a large rapidity gap adjacent to the proton 
have been observed in NC events with high $Q^2$ at HERA which are considered as
inelastic diffraction of the photon ({\it K. Borras; C. Kiesling}).
Similar events with the charged current have now been reported as well.
Assuming Regge factorization for small momentum transfer $t$ 
between the in and outgoing protons
this process can be described by Pomeron exchange where
the virtual photon scatters off the Pomeron which is 
emitted by the incoming proton with momentum fraction
$x_{\pom}=M_X^2/s$ where  $M_X$ denotes the hadronc mass of the 
$\gamma^* \pom$ 
and $\sqrt{s}$ the cms energy of the $\gamma^* p$ system. 
In analogy to the standard 
parton model for $ep$ DIS one can introduce  
Parton Distribution Functions for the Pomeron \cite{ischlein}
to describe $e \pom$ DIS. In this ``Diffractive DIS'' 
it is possible to derive QCD factorization 
of the photoabsorption cross section $\gamma^*p\to pX$ at fixed 
 $x_{\pom}$ and momentum transfer 
$t$ \cite{collins}
\begin{equation}
\frac{d^2\sigma^{DDIS}}{dx_{\pom} dt}\ =\ 
     \sum_q \int_x^{x_{\pom}} d\xi f_q^{D}(x_{\pom},t;\xi,Q^2)
     \sigma^{\gamma^*q}(x,Q^2,\xi)
 \label{Pfact}
\end{equation}
which further simplifies according to Regge factorization
$f_q^{D}(x_{\pom},t;
x,Q^2)=f_{\pom/p}(x_{\pom},t)f_{q/\pom}(\beta=x/x_{\pom},Q^2)$.
These PDF's are then studied as function of
$\beta=x_{q/\pom}$ or $\beta=x_{g/\pom}$  where Bjorken $x=\beta x_{\pom}$.
Both factorization properties are found to be satisfied by the data
but it is necessary to include Reggeon exchange in addition to 
Pomeron exchange. 

The very precise data from the full HERA-I analysis 
show the rise of the reduced 
cross section $\hat\sigma^{\gamma^*\pom}(\beta,Q^2)$
(suitably normalized to correspond to $F_2^{ep}$)
%(\ref{Pfact}) by the Pomeron flux $f_{\pom/p}(x_{\pom},t)$ and 
%some kinematic factors,  
with $Q^2$ for $\beta\lesssim 0.7$ and the decrease for higher $\beta$.
The change in slope occurs at much higher value than in case 
of $\gamma^*p$ scattering.
The striking pattern of scaling violation is explained in a NLO DGLAP fit 
(parameters  $\Lambda_{\overline{MS}}$, initial PDF's at $Q_0=3$ GeV)
by the large gluon fraction in the Pomeron 
$ f_g=75\pm15 \%. $
QCD factorization is also verified by the observation of diffractive 
di-jet and charm jet production in agreement with NLO QCD computations 
using the Pomeron PDF's so obtained.

Theoretical models for the diffractive structure functions ({\it J. Bartels})
have been developed
within the dipole picture with saturation, 
such as the models by GBW \cite{gbw} 
already mentioned and by BEKW \cite{bekw} which take 
the higher order QCD processes ($\gamma^*\to q\bar q g$) into acount
and provide a good description of $F_2^{DDIS}(\beta,Q^2)$.

The structure of simple Feynman diagrams for 
DDIS has been discussed by {\it S. Brodsky}. 
Explicit calculations in case of Feynman gauge for 
$\gamma^*q\to s\bar s q$
show that the rescattering of the struck $s$ quark 
involving nearly on-shell intermediate states
leads to an imaginary amplitude and an effective
Pomeron exchange in the production of the colour 
singlet $s\bar s$ state. This process survives in the Bjorken limit.
The same QCD final state interaction 
also can produce single spin asymmetries
in semi-inclusive DIS \cite{brodsky}.

\subsection{Diffraction in $p\bar p$ collisions}
Results from the TEVATRON on multi-gap events, 
hard diffractive  processes with jets, $W,Z,J/\psi,B$ have been discussed;
exclusive double Pomeron $\chi_c,\ \gamma\gamma$ and di-jet production
are of interest as benchmark for exclusive Higgs production at the LHC;
the analysis of RUN II data is in progress
(reports by {\it K. Borras, M. Convey, K. Goulianos}). 

The factorization of Pomeron processes has been established in DDIS 
but the arguments cannot be taken over to hadronic processes. In fact,
the observed cross sections for diffractive di-jet rates at the Tevatron,
are suppressed by an order of magnitude as expected 
using the Pomeron PDF's from HERA assuming factorization \cite{cdffacbd}. 
%
%%%%%%%%%%%%%% Begin Figure %%%%%%%%%%%%%%%%%%%%%%%%%%%%%%%%%%%%%%%%%%%%
\begin{figure*}
% \centering
\begin{center}
 \epsfig{file=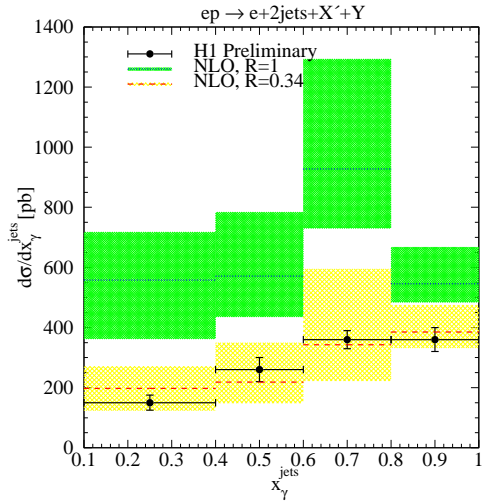,width=7cm}
\end{center}
 \caption{\label{f:diff} NLO cross sections for
 diffractive dijet photoproduction as functions of 
jet energy fraction $x_\gamma^{\rm jets}$
compared to preliminary H1 data. The predictions including absorption
(R=0.34 \protect\cite{kkmr}) agree with the data, 
those without absorption (R=1) do not (from Ref. \protect\cite{kk}).
% The shaded areas indicate a variation of scales by a factor of two around
% $E_T^{\rm jet1}$.}
\vspace{-0.5cm}
}
\end{figure*}
%%%%%%%%%%%%%% End of Figure %%%%%%%%%%%%%%%%%%%%%%%%%%%%%%%%%%%%%%%%%%%

Such a suppression has been expected from
reinteraction of spectator partons which yields a 
reduced gap survival probability, see for example \cite{bjgap}. 
This idea is supported by the observation of CDF that 
the occurence of an additional gap is unsuppressed;
by appropriate combination of 
multi-gap cross sections in soft diffraction the survival probability has been
determined as $S=0.23\pm 0.07$ at $\sqrt{s}=1800$ GeV. Taking this effect into 
account the factorization properties and agreement with extrapolations from 
HERA can be reestablished.
A general systematics of multi-gap events of various kinds
with characteristic factorization 
properties, also in hard processes, has been proposed
({\it K. Goulianos}) in the framework of a parton model 
which includes empirical rules
like ``$1/M^2$ scaling''  and ``Pomeron flux renormalization''.

The theoretical approach by KKMR \cite{kkmr} is based on hard QCD scattering 
processes, but includes initial state interaction by multiple Pomeron exchanges
which are derived in a 2-channel eikonal model. This approach explains 
quantitatively the
phenomena of factorization breakdown and the rates of multiple gap events.
An interesting prediction concerns the breakdown of 
factorization in di-jet photoproduction at HERA \cite{kkmr} with an additional 
suppression $S=0.34$. This effect has been verified recently in a NLO QCD 
calculation \cite{kk} in comparison with H1 data \cite{H1diff} 
(see Fig. \ref{f:diff}).  

\section{Heavy ion collsions and Evidence for Quark Gluon Plasma} 
Here the transition from a parton to a hadron ensemble is studied 
in a very high particle multiplicity environment 
which suggests a thermodynamic treatment. 
In lattice QCD one expects with increasing temperature
$T$ or energy density $\epsilon$ 
a phase transition from confined to deconfined matter, 
i.e. from a hadron gas to a 
Quark Gluon Plasma. The ratio $\epsilon/T^4$,
a measure of the number of degrees of freedom, 
 shows a characteristic
rise with $T$
near $T_0\sim 170$ MeV, $\epsilon_0\sim 0.7$ GeV/fm$^3$, 
depending also on the number of flavours,
over a range of about $\Delta T \sim 80$ MeV 
and then, above  $\epsilon \sim 2$ GeV/fm$^3$, 
becomes nearly $T$ independent; in this region there is
still a
large ($\sim 30$\%) deviation from the asymptotic
Stefan-Boltzmann limit for the ideal quark-gluon gas
\cite{karsch}. 
Another prediction concerns the dependence of 
the critical temperature $T_c$ on the baryochemical potential $\mu$ 
which can be tested through the hadron composition of the final state.

Previous research at the SPS has identified various signatures expected for
the transition to a QGP, such as strangeness excess, $J/\psi$-suppression, 
universal chemical freeze out; the energy density is found at 
$\epsilon\sim 2-4$ GeV/fm$^3$, 
just above the critical value. Now with RHIC a new regime with the much 
higher initial density of $\sim 15$ GeV/fm$^3$ is reached 
much above the critical density. This leads to new signatures:
a strong asymmetric flow of particles reflecting the initial spacial 
anisotropy and the
strong absorption of high $p_T$ jets in the nucleus 
(``jet quenching'').
It appears particularly impressive with the new RHIC data, that more specific
QCD tests are now becoming feasible. 
A general outline of the RHIC results
and their interpretation is given by {\it R. Seto}.

\subsection{Onset of deconfinement in the  SPS energy range}
Results on  $AA$ collisions 
from an energy scan over the lower SPS 
energies 20-80 AGeV  have been presented by {\it P. Seyboth}.
After the observation of signatures 
for QGP formation at the top SPS energy the aim was to search for 
an energy  threshold of such signatures.

A remarkable effect is seen in the energy dependence
of the freeze out temperature, as determined from the slope of the 
transverse mass of kaons: at low energies 
there is a strong rise of temperature 
followed by saturation over
the SPS energy range and continued rise at RHIC energies. 
This is the typical behaviour expected 
for a mixed phase of QGP and hadron gas 
at constant temperature.
Such a behaviour at the
energy density $\epsilon\gtrsim 2$ GeV/fm$^3$ in the SPS range 
matches the values
expected from lattice calculations.
A threshold effect is also seen for strangeness production, especially the 
$K^+/\pi^+$ ratio which is a well established signature for QGP formation.

Another test of lattice QCD calculations 
concerns the phase diagram in the variables 
$T$ vs. baryochemical potential $\mu_B$.
The analysis of  the particle species abundances in statistical models
yields values for $T$ and $\mu_B$ at freeze out which converge for low  
$\mu_B$ (high energies) towards the QCD expectation \cite{fodor}.

\subsection{Space-time evolution of collision process at RHIC}

{\it Initial stage (E.G. Ferreiro, K. Tuchin)}. The initial conditions 
can be introduced by the gluon density in the nucleus at the
saturation scale (see Eq. (\ref{qsat})) as 
$xG_A(x,Q_s^2)\sim\frac{\pi R_A^2Q_s^2(x,A)}{\alpha_s(Q_s^2)}$
(``Colour Glass Condensate'') where the $A$ dependence is inferred
from $G_A\sim A$, $\pi R_A^2\sim A^{\frac{2}{3}}$ and 
$Q_s^2\sim A^{\frac{1}{3}}$. Assuming proportionality of 
particle multiplicity and initial gluon rapidity density 
$dN/dy \sim xG(x,Q_s^2)$ one finds for an $AA$ collision with
$N_{part}$ participating nucleons 
a very slow increase of central hadron multiplicity 
$(1/N_{part}) dN/dy\sim 1/\alpha_s(Q^2_s)\sim  \ln N_{part}$
with energy and centrality \cite{kn}, a very successful prediction;
other predictions from parton saturation follow for the transverse momentum
distributions (review \cite{levin}).

{\it Early interactions: jet production and jet quenching (Miller, Vitev)}. 
These new phenomenona are related to the hard parton-parton scattering in
nuclear collisions and subsequent 
absorption of one parton in the dense medium. 
The absorption is mainly due to induced gluon 
radiative energy loss in multiple scattering inside the nucleus 
and is proportional to the 
plasma density \cite{quench}. The most striking evidence for jet quenching
comes from the study of azimuthal angle  correlations of particles 
associated
with a high $p_T$ trigger particle \cite{starquench}: whereas in both 
$pp$ and $dAu$ scattering one observes a jet in direction opposite to
the trigger jet
this away side jet is fully suppressed in central $AuAu$ collisions.
This is naturally explained by the hard collision taking 
place near the edge of the nucleus where one scattered 
parton leaves the nucleus undisturbed whereas
the second parton has to move through the big nucleus. This measurement
demonstrates the big difference between 
cold nuclear matter traversed in $dA$ collisions 
and the matter created in the central $AuAu$
collision. Theoretical calculations lead to an estimate of energy density
of $\epsilon \sim 15$ GeV/fm$^3$ corresponding to about 
100 times nuclear density. Additional studies confirm the 
absorption strength as function of the nuclear thickness in non-central 
collisions as well as the reappearence of the lost energy of the primary parton
in the soft particles in jet direction.

While these observations provide already 
a strong argument in favour of the presence of 
a QGP there are further crucial tests. QCD predicts 
the absorpion of a gluon to be stronger than that of the quark 
by the factor $C_A/C_F$. Furthermore,
heavy quarks are less absorbed
as the small angle radiation is cut off below $\Theta_c=m_Q/E$
(``dead cone effect'') \cite{dokHQ}. This effect is being searched for 
in nuclear D meson production ({\it Z. Xu}). 

 {\it Hydrodynamic evolution, flow phenomena
(Y. Hama, T. Hirano, H. Long, J. Velkowska and S. Voloshin)}. The 
hydrodynamics
description of the particle production process includes the initial condition
(energy density, initial flow), the EoS for the QGP 
with a phase transition built in (parameter: latent heat)  
and the freeze out mechanism for hadron production.
The observations related to hydrodynamics are:\\
1. Violation of $m_T$-scaling in $AA$ collisions, which denotes
 a universal slope $\beta$ 
in $dN/dp_T^2\sim e^{-\beta m_T}$  for particles of different mass
and works well in $pp$ scattering. In hydrodynamic flow the particles 
acquire similar velocities and therefore protons obtain higher momenta 
than pions.\\
2. An important consequence of hydrodynamics is the appearence of an asymmetric,
especially elliptic flow: in the non-central collision of two nuclei there is 
an almond shape overlap region which generates an asymmetric  pressure gradient
with maximum in impact direction; this results in a corresponding asymmetry
in energy and particle flow $\frac{dN}{d\phi}\sim 1+2v_2 \cos2\phi+ \ldots$
The elliptic flow $v_2$ increases for particles of higher $p_T$ with a delay 
for heavier particles.\\
3. The Equation of State (EoS) can be investigated, especially the
properties of the phase transition (latent heat 
$\sim 800$ MeV/fm$^3$) which provides another QCD test.\\
4. An apparent problem for the hydrodynamic description is met with the 
Bose-Einstein correlations between identical particles which depend on the
evolution of the space time volume containing the matter.
Whereas {\it T. Hirano} notes a failure of this description it has been 
pointed out by {\it T. Csorg\"o} in the discussion that the disagreement 
can be avoided by a proper choice of the initial transverse flow
to explain the final ``Hubble flow'' (Buda-Lund model \cite{buda-l}). 

\begin{figure*}
 \centering   
 \epsfig{file=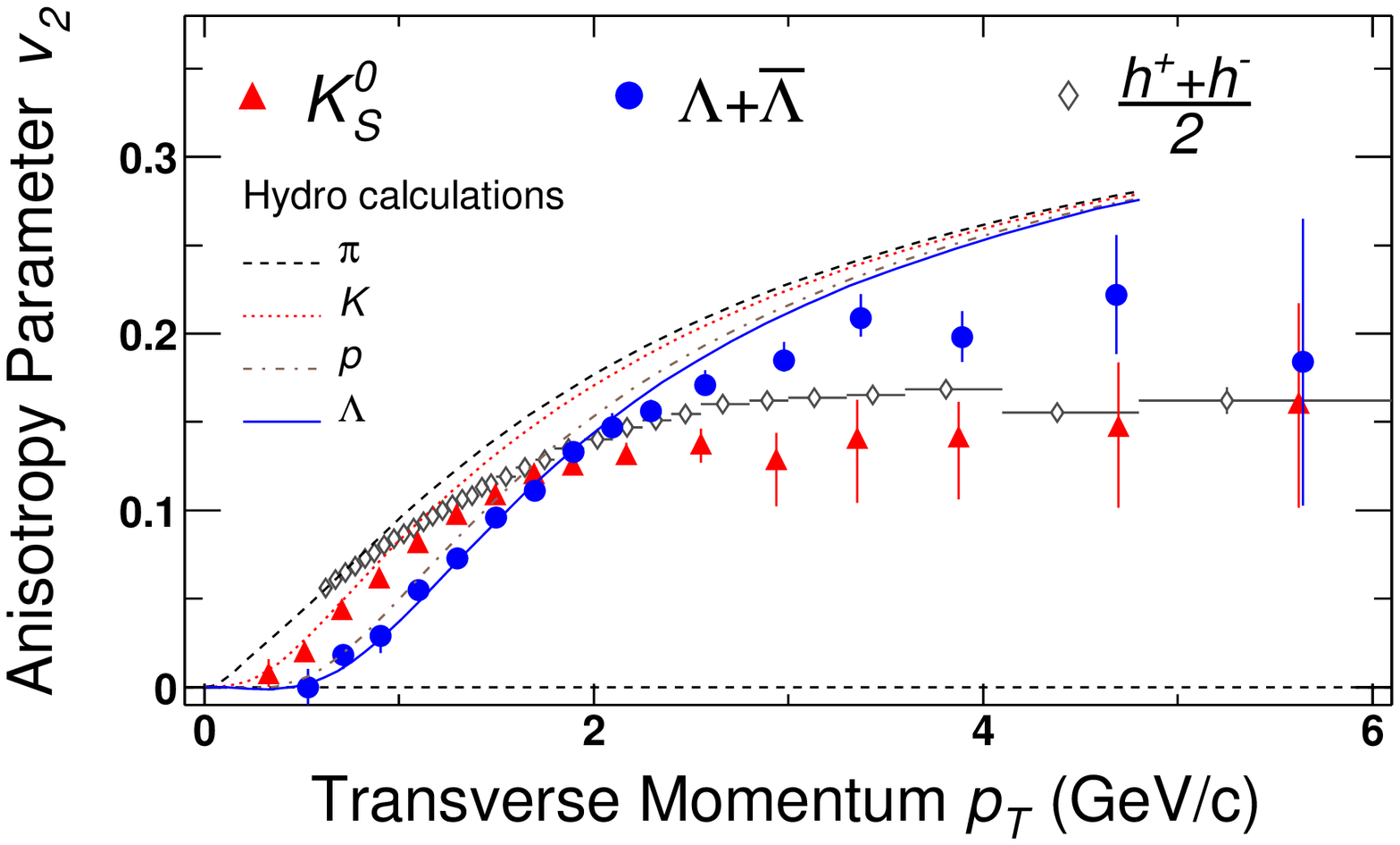,width=0.49\textwidth}
 \epsfig{file=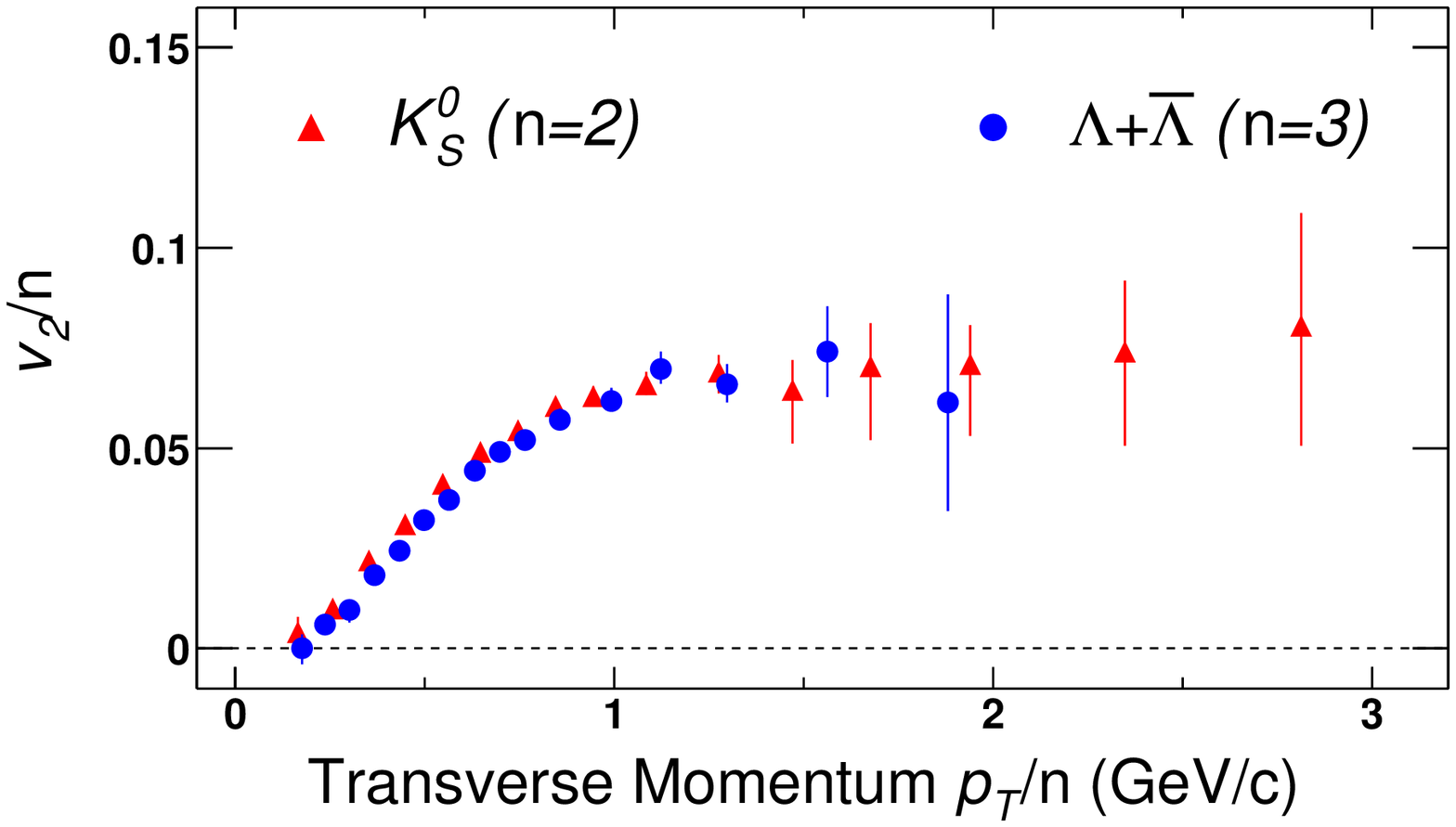,width=0.49\textwidth}
 \caption{\label{fig.k0lambda} 
Transverse momentum dependence of 
elliptic flow $v_2$ measured by STAR \protect\cite{starval}
a) before and b) after rescaling by the number of valence quarks $n$.
\vspace{-0.5cm}
\label{fig:val}
}
\end{figure*}
{\it Coalescence, quark recombination (R. Hwa, J. Velkovska, S. Voloshin)}. 
Another striking new phenomenon observed at RHIC
is the grouping of spectra according to the number of constituent quarks.
This is observed in the $p_T$ dependence of 
the elliptic flow parameter $v_2$ ($p_T<6$ GeV) where $\pi,K$ and 
$p,\Lambda,\Xi$ fall into separate bands but show a uniform dependence
if rescaled according to the number $n_V$ of valence quarks 
\cite{starval,phenixval}
\begin{equation} 
v_2/n_V=f(p_T/n_V).
\label{valence}
\end{equation}
As an example the $K,\Lambda$ spectra are shown in Fig. \ref{fig:val}.
Another observation concerns the ratios $R_{CP}$ of particle spectra 
for central and peripheral collisions where mesons 
($\phi,K^0, K^\pm$) and baryons
 ($\Omega,\Xi,\Lambda +$ antiparticles) fall into separate bands
 in the region 
$2\lesssim p_T\lesssim 6$ GeV. This confirms the idea of parton coalescence
\cite{mv}. There are several other ``anomalies'' in nuclear production,
for example, the large ratio $p/\pi^+$ for large $p_T>2$ GeV which can be 
explained within a recombination mechanism for thermal and shower partons
({\it R. Hwa}). 

\begin{figure}[t!]%[hbt]
\begin{center}
\vspace{-3.1cm}
\includegraphics[width=10cm]{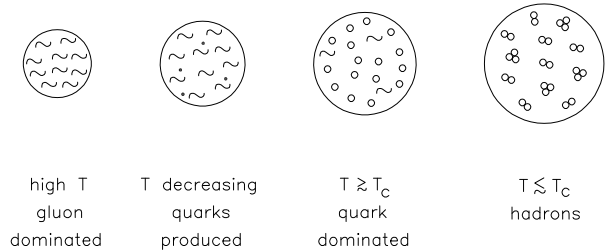}
\end{center}
\vspace{-7.5cm}
%\vspace{-11.0cm}
\caption{
Evolution of the quark gluon plasma from a gluon dominated  phase of 
high temperature and density towards a quark dominated phase near the critical 
temperature $T_c$, finally the transition 
from constituent quarks to hadrons.
}
\label{evol}
\end{figure}

The behaviour (\ref{valence}) suggests that before hadron formation
there is a flow of constituent quarks which then recombine 
into the observed hadrons.
This implies a strong dependence
 of the $q/g$ composition of the plasma during the evolution 
as illustrated in Fig. \ref{evol} : initially the primary collision produces
mainly gluons at high temperature; during the expansion $q\bar q$ 
pairs are produced, but in approaching the critical temperature the 
gluons are absorbed by the quarks 
(``constituent quarks'') which then by coalescence form the final state 
hadrons. The strong ordering according to valence content
is against expectations from a hadronic resonance gas which would group 
particles according to mass. These observations
are therefore another strong argument for the presence of a QGP (although
at the end without gluons). Looked at in reverse order the evolution depicted
 in Fig. \ref{evol} is quite natural: 
the hadrons under increased pressure dissolve at first 
into constituent quarks (as in the additive quark model) but under 
increased pressure gluons are easily freed from the constituent quarks
and yield a genuine QGP.

{\it Strongly interacting QGP}. There is another interesting message in the
$p_T$-dependence of the elliptic flow parameter $v_2$. Using a calculation
for a QGP within transport theory \cite{gmtransport} it is found that the 
gluon-gluon cross sections of few mb expected in pQCD would give negligeable 
flow effects, only cross sections of $\gtrsim 40$ mb would yield the observed 
asymmetric flow. Therefore, the data suggest a strongly interacting QGP
and this can be related to the large deviation from the Stefan-Boltzmann limit
of the ideal gas found in the lattice QCD calculations \cite{shuryak}.

\section{Other presentations}
Finally, we list a few other topics which have been discussed.\\
{\it 1. Hadronic Phenomena}.
Particle correlations are often not accessible to a QCD description, 
especially the Bose-Einstein correlations 
which turn out to be particularly important in the
discussion of Heavy Ion Collisions. Also there are discussions of critical
phenomena such as percolation and clustering in a string or hadron model 
framework.\\
{\it 2. Hadron Spectroscopy}. This is
another active field with surprising results on unexpected hadronic
particles, including the still controversial ``pentaquarks'', especially
$\Theta^+(1540)$, not observed at TEVATRON, and the new charmonium state
with decay
$X(3872)\to J/\psi \pi^+\pi^-$ reported 
here by CDF, which stimulates discussions of nonperturbative aspects of QCD.\\
{\it 3. Astroparticle Physics}.
In this field we witness a flourishing activity where Multiparticle
Production plays an important role, although not as the basic goal of the
activity but rather as a tool. Primary cosmic particles of
superhigh energies ($10^8$ TeV) are studied. The interpretation 
of the particle yields,
in particular the determination of the primary particle energies requires a
detailed understanding of the propagation and showering of the cosmic rays
(elementary particles or nuclei) in various media (air, earth, ice),
including the saturation
phenomena and production and decay of heavy quarks ($c,b$ quarks).
Questions of the role of particles in theories beyond the
standard model are being discussed, as well as the origin of such high
energy cosmic rays.  
%The Quark gluon plasma in a thermal environment should
%be formed in the early universe before $10^{-6}$ sec.

\section{Summary}
The number of multiparticle production 
phenomena which can be explained within QCD is steadily increasing.\\
{\it 1. Hard processes:} Calculations for
larger kinematical ranges, for observables of higher complexity
 and with increasing accuracy agree with the data, 
no definitive failures have been reported this time.\\
{\it 2. Soft particle production:} It follows perturbative QCD
expectations surprisingly well which is in support of a parton hadron duality picture 
of hadronization with soft colour confinement.\\
{\it 3. Small $x$ and diffraction:} For a high density 
regime at small $Q^2$ there is indirect evidence for saturation 
(``Color Glass Condensate'') inferred from the success of saturation models.
Intrinsic Pomeron structure is a useful concept in diffractive scattering,
the systematics of factorization and its breaking are becoming better 
understood.\\
{\it 4. Heavy Ions and QGP:} 
%There is a dramatic development from the more
%phenomenological studies at the SPS towards genuine QCD tests at RHIC.
There is an indication of a phase transition with a mixed phase 
over the SPS energy range.
The higher initial pressure and longer evolution time available at RHIC 
have provided clear evidence for jet quenching and strong elliptic flow
with an initial energy density about 100 times higher 
than in nuclear matter and an order of magnitude above critical density.
These phenomena are adequately 
described in terms of a strongly interacting QGP;
these observables can be used as new diagnostic tools which allow
detailed tests of (perturbative and non-perturbative) QCD predictions: 
parton type dependence of absorpion, 
Equation of State with latent heat and strong deviation from ideal gas limit,
phase diagram $T_c$ vs. $\mu_B$, the initial CGC state. 
Hadrons are formed by coalescence of constituent quarks which dominate the
QGP in its final stage.

\section*{Acknowledgement}
I would like to thank Bill Gary and his crew for their engagement in 
organizing this lively meeting
which managed to bring together and to mix up 
the different multiparticle communities to the benefit of all of us.
I am also grateful for the helpful discussions 
with participants of the meeting,
especially J. Bartels, V. Khoze, C. Kiesling, N. Schmitz and P. Seyboth.

\end{document}